\makeatletter
\newcommand{\dontusepackage}[2][]{%
  \@namedef{ver@#2.sty}{9999/12/31}%
  \@namedef{opt@#2.sty}{#1}}
\makeatother
\dontusepackage{subfigure}

\documentclass[]{article}

\usepackage{lmodern}
\usepackage{amssymb,amsmath}
\usepackage{ifxetex,ifluatex}
\usepackage[usenames,dvipsnames]{color}
\usepackage{fixltx2e} 
\ifnum 0\ifxetex 1\fi\ifluatex 1\fi=0 
  \usepackage[T1]{fontenc}
  \usepackage[utf8]{inputenc}
\else 
  \ifxetex
    \usepackage{mathspec}
    \usepackage{xltxtra,xunicode}
  \else
    \usepackage{fontspec}
  \fi
  \defaultfontfeatures{Mapping=tex-text,Scale=MatchLowercase}
  
\fi
\IfFileExists{upquote.sty}{\usepackage{upquote}}{}
\IfFileExists{microtype.sty}{%
\usepackage{microtype}
\UseMicrotypeSet[protrusion]{basicmath} 
}{}
\usepackage[margin=1.0in,bottom=1.5in]{geometry}
\usepackage[square,numbers,sort&compress]{natbib}
\usepackage{listings}
\usepackage{graphicx}
\makeatletter
\def\maxwidth{\ifdim\Gin@nat@width>\linewidth\linewidth\else\Gin@nat@width\fi}
\def\maxheight{\ifdim\Gin@nat@height>\textheight\textheight\else\Gin@nat@height\fi}
\makeatother
\setkeys{Gin}{width=\maxwidth,height=\maxheight,keepaspectratio}
\usepackage{caption}
\usepackage{float}
\usepackage{subfig}
\usepackage{algorithm} 
\let\scholmdAlgorithm\algorithm
\let\endscholmdAlgorithm\endalgorithm
\let\algorithm\relax \let\endalgorithm\relax

{
 \catcode`\*=11\relax
 \global\let\scholmdAlgorithm*\algorithm*
 \global\let\endscholmdAlgorithm*\endalgorithm*
 \global\let\algorithm*\relax 
 \global\let\endalgorithm*\relax
}
\ifxetex
  \usepackage[setpagesize=false, 
              unicode=false, 
              xetex]{hyperref}
\else
  \usepackage[unicode=true]{hyperref}
\fi
\hypersetup{breaklinks=true,
            bookmarks=true,
            pdfauthor={},
            pdftitle={Weak deep priors for seismic imaging},
            colorlinks=true,
            citecolor=black,
            urlcolor=blue,
            linkcolor=black,
            pdfborder={0 0 0}}
\usepackage[preprint]{neurips_2019}
\usepackage{url}
\usepackage{booktabs}
\usepackage{amsfonts}
\usepackage{nicefrac}
\usepackage{microtype}

\title{Weak deep priors for seismic imaging}
\author{Ali Siahkoohi, Gabrio Rizzuti, and Felix J. Herrmann\\School of
Computational Science and Engineering,\\Georgia Institute of
Technology\\\texttt{\{alisk,\phantom{\ }rizzuti.gabrio,\phantom{\ }felix.herrmann\}@gatech.edu}}
\date{}

\begin{document}
\maketitle
\begin{abstract}
Incorporating prior knowledge on model unknowns of interest is essential
when dealing with ill-posed inverse problems due to the nonuniqueness of
the solution and data noise. Unfortunately, it is not trivial to fully
describe our priors in a convenient and analytical way. Parameterizing
the unknowns with a convolutional neural network (CNN), and assuming an
uninformative Gaussian prior on its weights, leads to a variational
prior on the output space that favors ``natural'' images and excludes
noisy artifacts, as long as overfitting is prevented. This is the
so-called deep-prior approach. In seismic imaging, however, evaluating
the forward operator is computationally expensive, and training a
randomly initialized CNN becomes infeasible. We propose, instead, a weak
version of deep priors, which consists of relaxing the requirement that
reflectivity models must lie in the network range, and letting the
unknowns deviate from the network output according to a Gaussian
distribution. Finally, we jointly solve for the reflectivity model and
CNN weights. The chief advantage of this approach is that the updates
for the CNN weights do not involve the modeling operator, and become
relatively cheap. Our synthetic numerical experiments demonstrate that
the weak deep prior is more robust with respect to noise than
conventional least-squares imaging approaches, with roughly twice the
computational cost of reverse-time migration, which is the affordable
computational budget in large-scale imaging problems.
\end{abstract}

\section{Introduction}\label{introduction}

Linearized seismic imaging involves an inconsistent, ill-conditioned
linear inverse problem due to presence of shadow zones and complex
structures in the subsurface, coherent linearization errors, and noisy
data. Due to nonuniqueness, using prior information as regularization is
essential. This particular choice is crucial because it typically
affects the final result. Conventional methods mostly rely on
handcrafted and unrealistic priors, such as a Gaussian or Laplace
distributed model parameters (in the physical or in a transform domain).
These simplifying assumptions, while being practical, negatively bias
the outcome of the inversion.

Recent proposals
\citep{Lempitsky, Cheng_2019_CVPR, gadelha2019shape, liu2019deep, wu2019parametric, shi2020deep, siahkoohi2020EAGEdlb}
make use of convolutional neural networks (CNN) as a prior.
Specifically, \citet{siahkoohi2020EAGEdlb} reparameterize the unknown
reflectivity model by a CNN and impose a Gaussian prior on its weights.
These authors show that the combination of the functional form of a CNN
and a Gaussian prior on its weights is a suitable prior for seismic
imaging. However, since every update to CNN weights requires the action
of the forward operator and its adjoint, tuning randomly initialized CNN
weights need many stochastic optimization steps. In seismic imaging,
computing the action of the forward operator---i.e., linearized Born
scattering operator, and its adjoint is computationally expensive, which
might limit the application of deep priors.

We propose the \emph{weak deep prior}, a computationally convenient
formulation that relaxes deep priors. Instead of reparameterizing the
unknowns with CNNs, we let the unknown reflectivity to be distributed
according to a Gaussian distribution centered at the CNN network output.
Next, we jointly solve for the reflectivity model and CNN weights. This
formulation decouples the forward operator with the CNN, allowing for
fast and forward-operator free updates of CNN weights, while partially
keeping the advantages of the deep prior. The proposed formulation
additionally allows for imposing handcrafted or physical hard
constraints on the unknowns, which is often not feasible when imposing
deep priors \citep{herrmann2019NIPSliwcuc}.

In general, numerous efforts involve the incorporation of ideas from
deep learning in seismic processing and inversion
\citep{ovcharenko2019deep, rizzuti2019EAGElis, siahkoohi2019dlwr, siahkoohi2019transfer, siahkoohi2019srmedl, sun2019extrapolated, zhang2019regularized}.
Deep prior itself have been utilized by \citet{liu2019deep} to perform
seismic data reconstruction. \citet{wu2019parametric} propose to
pretrain a randomly initialized CNN before reparameterizing the velocity
model in the context of Full-Waveform Inversion. \citet{shi2020deep} use
the deep priors in the context of denoising. Finally,
\citet{siahkoohi2020EAGEdlb} proposes a deep-prior based Bayesian
framework for seismic imaging and perform uncertainty quantification.

Our work is organized as follows. We first introduce the original
concept of deep prior and how it can be integrated in seismic imaging.
Next, we develop the weak deep prior framework and the associated
optimization problem. We conclude by showcasing the proposed method
using a synthetic example involving a 2D portion of a real migrated
image of the \href{https://wiki.seg.org/wiki/Parihaka-3D}{3D Parihaka}
dataset \citep{Veritas2005, WesternGeco2012} in the presence of strong
noise.

\section{Seismic imaging}\label{seismic-imaging}

Seismic imaging is the problem of estimating the short-wavelength
structure of the Earth's subsurface, denoted by $\delta \mathbf{m}$
given data recorded at the surface,
$\delta \mathbf{d}_{i}, \ i = 1,2, \cdots , N$, where $N$ is the number
of shot records. Besides observed data, this inverse problem requires a
smooth background squared-slowness model, $\mathbf{m}_0$, and estimated
source signatures, $\mathbf{q}_i$. When noise in the data can be
approximated by a zero-mean Gaussian random variable, $\ell_2$-norm data
discrepancy defines the likelihood function
\citep{tarantola2005inverse}. Assuming the noise covariance is
$\sigma^2 \mathbf{I}$, we can write the negative log-likelihood of the
observed data as follows:
\begin{equation}
\begin{aligned}
&\ - \log p_{\text{like}} \left ( \left \{ \delta \mathbf{d}_{i} \right \}_{i=1}^N \normalsize{|}\delta \mathbf{m} \right )   = -\sum_{i=1}^N  \log p_{\text{like}} \left ( \delta \mathbf{d}_{i}  
 \normalsize{|}\delta \mathbf{m} \right ) \\
&\ = \frac{1}{2 \sigma^2} \sum_{i=1}^N  \|\delta \mathbf{d}_i- \mathbf{J}(\mathbf{m}_0, \mathbf{q}_i) \delta \mathbf{m}\|_2^2 \quad + \underbrace {\text{const}}_{\text{Ind. of }  \delta \mathbf{m}}. \\
\end{aligned}
\label{nll}
\end{equation}
 In these expressions, $p_{\text{like}}$ denotes the likelihood
probability density function, and $\mathbf{J}$ is the linearized Born
scattering operator. The maximum likelihood estimate (MLE), denoted by
$\widehat{\delta \mathbf{m}}_{\text{MLE}}$, is obtained by minimizing
the negative-log likelihood defined in Equation~\ref{nll} with respect
to $\delta \mathbf{m}$. Notoriously, MLE estimators tend to produce
imaging artifacts. To address this issue, we discuss a special kind of
prior based on neural networks: the so-called deep priors.

\section{Imaging with deep priors}\label{imaging-with-deep-priors}

Parameterizing the unknown variables with a CNN, with a fixed input, has
shown promising results in inverse problems
\citep{Lempitsky, Cheng_2019_CVPR, gadelha2019shape, liu2019deep, wu2019parametric, shi2020deep, siahkoohi2020EAGEdlb}.
In this approach, weights and biases are Gaussian random variables and
they are tuned to fit the observed data. The success of this approach
hinges on the special structure of the CNN, which tends to favor
noise-free looking images. Despite this feature, it should be noted that
a stopping criteria is still essential to avoid overfitting the noise in
observed data. Notwithstanding this challenge, we propose to
parameterize the unknown reflectivity model by a CNN---i.e.,
$\delta \mathbf{m} = {g} (\mathbf{z}, \mathbf{w})$, where
$\mathbf{z} \sim \mathrm{N}( \mathbf{0}, \mathbf{I})$ is the fixed input
to the CNN and $\mathbf{w}$ denotes the unknown CNN weights. Imposing a
Gaussian prior on $\mathbf{w}$ with covariance matrix
$\lambda^{-2}\mathbf{I}$ allows us to formulate the negative
log-posterior distribution for $\mathbf{w}$ as follows:
\begin{equation}
\begin{aligned}
&\ p_{\text{post}} \left ( \mathbf{w} \normalsize{|}  \left \{ \delta \mathbf{d}_{i} \right \}_{i=1}^N \right ) \propto  \left [ \prod_{i=1}^{N} p_{\text{like}} \left ( \delta \mathbf{d}_{i}  \normalsize{|}\mathbf{w}  \right ) \right ]   p_{w} \left ( \mathbf{w} \right ), \\
&\ \text{where} \quad  p_{w} \left ( \mathbf{w} \right ) = \mathrm{N}(\mathbf{w} \normalsize{|} \mathbf{0}, \lambda^{-2}\mathbf{I}). \\
\end{aligned}
\label{deep-prior}
\end{equation}
 In the equation above, $p_{w}$ and $p_{\text{post}}$ denote the prior
and posterior probability density functions, respectively. The maximum a
posteriori estimator (MAP), denoted by
$\widehat{\mathbf{w}}_{\text{deep}}$, is obtained by maximizing
Equation~\ref{deep-prior} with respect to $\mathbf{w}$.

As stated before, there are two challenges in employing deep priors in
seismic imaging. The first challenge is finding a stopping criteria
while maximizing the posterior in Equation~\ref{deep-prior} to prevent
noise overfit. \citet{siahkoohi2020EAGEdlb} propose to perform
stochastic gradient Langevin dynamics
\citep[SGLD,][]{welling2011bayesian} steps to obtain samples from this
posterior distribution. Using these samples, these authors approximate
the conditional mean estimator, which prevents overfitting and at the
same time, yields a seismic image that has less imaging artifacts
compared to the MAP estimator. However, sampling the posterior is a
challenging feat in and of itself and is outside of the scope of this
discussion. Another challenge associated with deep-prior based imaging
is the number of iterations needed to optimize the CNN weights. Unless
the CNN is pretrained, its weights are initialized randomly, hence,
solving for $\mathbf{w}$ requires many iterations involving the seismic
modeling operator and its adjoint and may not be computationally
practical. Unfortunately, unlike other imaging modalities, such as
medical imaging, we generally do not have access to detailed information
on the subsurface. This limits the scope of the pretraining phase, which
in turn might adversely bias the outcome of the inversion, and
contradicts the premises of this work. In the next section, we introduce
our proposed method and discuss how to address the computational
challenges associated with optimizing the CNN's randomly initialized
weights, while keeping the advantages of the deep-prior based imaging.

\section{Imaging with weak deep
prior}\label{imaging-with-weak-deep-prior}

The deep-prior based imaging problem can equivalently be casted as the
following constrained optimization problem:
\begin{equation}
\begin{aligned}
&\ \mathop{\rm arg\,min}_{\delta \mathbf{m},\, \mathbf{w}} \left [ \frac{1}{2 \sigma^2} \sum_{i=1}^N  \|\delta \mathbf{d}_i- \mathbf{J}(\mathbf{m}_0, \mathbf{q}_i) \delta \mathbf{m} \|_2^2  +  \frac{\lambda^2}{2} \| \mathbf{w} \|_2^2   \right ] \\
&\  \text{subject to}\quad \delta \mathbf{m}=g (\mathbf{z}, \mathbf{w}), \\
\end{aligned}
\label{nlp-deep-prior}
\end{equation}
 where we restrict the feasible model to the output of
$g (\mathbf{z}, \mathbf{w})$. To address the computational challenge
associated with deep-prior based imaging, we propose to relax the
constraint in problem~\ref{nlp-deep-prior} and let $\delta \mathbf{m}$
be a random variable distributed according to a Gaussian distribution
centered at $g (\mathbf{z}, \mathbf{w})$ with covariance matrix
$\gamma^{-2}\mathbf{I}$. We denote the defined prior on
$\delta \mathbf{m}$ as the \emph{weak} deep prior. By decoupling the
forward operator and the CNN weights, observed data becomes
conditionally independent from $\mathbf{w}$, given $\delta \mathbf{m}$.
We can write the joint posterior distribution for
$(\delta \mathbf{m},\mathbf{w})$ using the defined prior as follows:
\begin{equation}
\begin{aligned}
&\ p_{\text{post}} \left (\delta \mathbf{m}, \mathbf{w} \normalsize{|} \left \{ \delta \mathbf{d}_{i} \right \}_{i=1}^N \right ) \\
&\ \propto \left [ \prod_{i=1}^{N} p_{\text{like}} \left ( \delta \mathbf{d}_{i}  \normalsize{|}\delta \mathbf{m}  \right ) \right ] p_{\mathbf{\text{weak}}} \left ( \delta \mathbf{m} \normalsize{|} \mathbf{w} \right )p_{w}(\mathbf{w}), \\
&\ \text{where} \quad p_{\mathbf{\text{weak}}} \left ( \delta \mathbf{m} \normalsize{|} \mathbf{w} \right ) = \mathrm{N}( \delta \mathbf{m} \normalsize{|} g (\mathbf{z}, \mathbf{w}), \gamma^{-2}\mathbf{I}). \\
\end{aligned}
\label{weak-deep-prior}
\end{equation}
 In Equation~\ref{weak-deep-prior},
$p_{\mathbf{\text{weak}}} \left ( \delta \mathbf{m} \normalsize{|} \mathbf{w} \right )$
denotes the weak deep prior, which is equivalent to a Gaussian
distribution centered at $g (\mathbf{z}, \mathbf{w})$ with covariance
matrix $\gamma^{-2}\mathbf{I}$. $\gamma$ is a hyperparameter that needs
to be tuned. We solve the imaging with weak deep prior problem by
minimizing the negative log-posterior defined in
Equation~\ref{weak-deep-prior} as follows:
\begin{equation}
\begin{aligned}
\widehat{\delta \mathbf{m}}_{\text{weak}}, \widehat{\mathbf{w}}_{\text{weak}} =\mathop{\rm arg\,min}_{\delta \mathbf{m},\, \mathbf{w}} &\ \left [ \frac{1}{2 \sigma^2} \sum_{i=1}^N  \|\delta \mathbf{d}_i- \mathbf{J}(\mathbf{m}_0, \mathbf{q}_i) \delta \mathbf{m} \|_2^2  \right. \\
&\ \left. \ +  \ \frac{\gamma^2}{2} \| \delta \mathbf{m}  - g (\mathbf{z}, \mathbf{w}) \|_2^2 + \frac{\lambda^2}{2} \| \mathbf{w} \|_2^2 \vphantom{\sum_{i=1}^N} \right ] \\
\end{aligned}
\label{nlp-weak-deep-prior}
\end{equation}
 where $\widehat{\delta \mathbf{m}}_{\text{weak}}$ and
$\widehat{\mathbf{w}}_{\text{weak}}$ are the obtained reflectivity and
CNN weights by solving the imaging with weak deep prior problem. We
consider $\widehat{\delta \mathbf{m}}_{\text{weak}}$ as the final
estimate in this approach. When $\gamma \rightarrow \infty$, the
solution to problem~\ref{nlp-weak-deep-prior} is the same as the
solution to problem~\ref{nlp-deep-prior}.

In formulation above, updating the parameters $\mathbf{w}$ does not
involve the action of the forward operator, hence, weights of the CNN
can be quickly and independently updated. Moreover, the optimization
problem~\ref{nlp-weak-deep-prior} offers flexibility to impose any
intersection of physical or handcrafted hard constraints, $\mathcal{C}$,
by limiting the search space to $\delta \mathbf{m} \in \mathcal{C}$
while minimizing the objective with respect to $\delta \mathbf{m}$,
using standard constrained optimization techniques
\citep{peters2018pmf}. In a similar fashion,
\citet{herrmann2019NIPSliwcuc} use a Total-Variation constraint in the
context of seismic imaging to jointly solve the imaging problem and
train a generative model capable of directly sampling the posterior
using the Expectation-Maximization method. As the main contribution of
this work, we choose not to utilize hard constraints and focus on the
computational aspect of the weak deep prior.

\section{Algorithm and implementation
details}\label{algorithm-and-implementation-details}

To limit the computational cost---i.e., number of wave-equation solves,
we use stochastic optimization algorithms to solve the optimization
problems~\ref{nlp-deep-prior} and~\ref{nlp-weak-deep-prior}. We
approximate the negative-log likelihood term (see Equation~\ref{nll})
using a single simultaneous source, made of a Gaussian weighted source
aggregate. While we could use stochastic gradient descent algorithm
\citep[SGD,][]{Robbins2007ASA}, we avoid it because of several
challenges associated with it. For example, even though SGD's ``noisy''
(approximate) gradient is an unbiased estimate of true gradient, its
variance is proportional to square of the step size. Therefore, choosing
the step size is a trade-off between convergence speed and accuracy.
Additionally, SGD updates different components of the unknown with the
same step size---i.e., no preconditioning, which is not desirable when
the objective has varying sensitivity with respect to different
components of the unknowns. Various stochastic optimization algorithms
to some extent address these issues by diagonally weighting the gradient
by the norm of the past gradients \citep{adagrad} or the (weighted) mean
of past squared gradients \citep{rmsprop}. We use Adagrad
\citep{adagrad} with step size $2 \times 10^{-3}$ to update
$\delta \mathbf{m}$ while estimating
$\widehat{\delta \mathbf{m}}_{\text{MLE}}$, and when solving
optimization problem~\ref{nlp-weak-deep-prior}. To update $\mathbf{w}$,
either in optimization problem~\ref{nlp-deep-prior}
or~\ref{nlp-weak-deep-prior}, we use RMSprop \citep{rmsprop} with step
size $10^{-3}$. We set the step sizes by extensive hyper-parameter
tuning. In Algorithm~\ref{alg}, which summarizes our proposed approach,
\texttt{Adagrad} and \texttt{RMSprop} are optimization subroutines that
given the objective value and the step size, provide an update for
$\delta \mathbf{m}$ and $\mathbf{w}$, respectively.

\begin{scholmdAlgorithm}
\textbf{Input:}\\\hspace*{0.333em}\hspace*{0.333em}\hspace*{0.333em}\hspace*{0.333em}\hspace*{0.333em}\hspace*{0.333em}$\mathbf{z} \sim \mathrm{N}( \mathbf{0}, \mathbf{I})$:~fixed~input~to~the~CNN\\\hspace*{0.333em}\hspace*{0.333em}\hspace*{0.333em}\hspace*{0.333em}\hspace*{0.333em}\hspace*{0.333em}$\lambda, \ \gamma$:~trade-off~parameters\\\hspace*{0.333em}\hspace*{0.333em}\hspace*{0.333em}\hspace*{0.333em}\hspace*{0.333em}\hspace*{0.333em}$\sigma^2$:~estimated~noise~variance\\\hspace*{0.333em}\hspace*{0.333em}\hspace*{0.333em}\hspace*{0.333em}\hspace*{0.333em}\hspace*{0.333em}$T$:~stochastic~optimization~steps~for~$\delta \mathbf{m}$\\\hspace*{0.333em}\hspace*{0.333em}\hspace*{0.333em}\hspace*{0.333em}\hspace*{0.333em}\hspace*{0.333em}$K$:~inner~loop~stochastic~optimization~steps~for~$\mathbf{w}$\\\hspace*{0.333em}\hspace*{0.333em}\hspace*{0.333em}\hspace*{0.333em}\hspace*{0.333em}\hspace*{0.333em}$\eta$,~$\tau$:~step~sizes~to~update~$\delta \mathbf{m}$~and~$\mathbf{w}$,~respectively\\\hspace*{0.333em}\hspace*{0.333em}\hspace*{0.333em}\hspace*{0.333em}\hspace*{0.333em}\hspace*{0.333em}$\left \{ \delta \mathbf{d}_{i}, \delta \mathbf{q}_{i} \right \}_{i=1}^N$:~observed~data~and~source~signatures\\\hspace*{0.333em}\hspace*{0.333em}\hspace*{0.333em}\hspace*{0.333em}\hspace*{0.333em}\hspace*{0.333em}$\mathbf{m}_0$:~smooth~background~squared-slowness~model\\\hspace*{0.333em}\hspace*{0.333em}\hspace*{0.333em}\hspace*{0.333em}\hspace*{0.333em}\hspace*{0.333em}\texttt{Adagrad}:~Adagrad~algorithm~to~update~$\delta \mathbf{m}$\\\hspace*{0.333em}\hspace*{0.333em}\hspace*{0.333em}\hspace*{0.333em}\hspace*{0.333em}\hspace*{0.333em}\texttt{RMSprop}:~RMSprop~algorithm~to~update~$\mathbf{w}$\\\textbf{Initialization:}\\\hspace*{0.333em}\hspace*{0.333em}\hspace*{0.333em}\hspace*{0.333em}\hspace*{0.333em}\hspace*{0.333em}Randomly~initialize~CNN~parameters,~$\mathbf{w}$\\\hspace*{0.333em}\hspace*{0.333em}\hspace*{0.333em}\hspace*{0.333em}\hspace*{0.333em}\hspace*{0.333em}$\delta \mathbf{m} = \mathbf{0}$\\1.~\textbf{for}~$t=1$~\textbf{to}~$T$~\textbf{do}\\2.~~~~Randomly~sample~$(\delta \mathbf{d}, \mathbf{q})$~from~$\left \{ \delta \mathbf{d}_{i}, \mathbf{q}_{i} \right \}_{i=1}^N$\\3.~~~~$\mathcal{L}(\delta \mathbf{m}) = \frac{N}{2 \sigma^2} \|\delta \mathbf{d}- \mathbf{J}(\mathbf{m}_0, \mathbf{q}) \delta \mathbf{m} \|_2^2 + \frac{\gamma^2}{2} \| \delta \mathbf{m} - g (\mathbf{z}, \mathbf{w}) \|_2^2$\\4.~~~~$\delta \mathbf{m} \leftarrow$~\texttt{Adagrad}~$(\mathcal{L}(\delta \mathbf{m}), \eta)$\\5.~~~~\textbf{for}~$k=1$~\textbf{to}~$K$~\textbf{do}\\6.~~~~~~~~$\mathcal{L}(\mathbf{w})= \frac{\gamma^2}{2} \| \delta \mathbf{m} - g (\mathbf{z}, \mathbf{w}) \|_2^2+ \frac{\lambda^2}{2} \| \mathbf{w} \|_2^2$\\7.~~~~~~~~$\mathbf{w} \leftarrow$~\texttt{RMSprop}~$(\mathcal{L}(\mathbf{w}), \tau)$\\8.~~~~\textbf{end~for}\\9.~\textbf{end~for}\\\textbf{Output:}~$\delta \mathbf{m}$
\caption{Seismic imaging with weak deep prior.}\label{alg}
\end{scholmdAlgorithm}

As mentioned before, the weak deep prior allows for fast updates of the
CNN weights (see the inner loop in lines $5$ -- $8$ of
Algorithm~\ref{alg}). However, choosing the number of updates for
$\mathbf{w}$ per each $\delta \mathbf{m}$ update is a trade-off between
reducing computational cost (many $\mathbf{w}$ updates) and preserving
the the deep prior advantages (maintained by employing several
$\mathbf{w}$ updates). In the extreme case, if we update $\mathbf{w}$
once per $\delta \mathbf{m}$ update, there is no computational gain
compared to the deep-prior based approach. On the other hand, if we
solve for $\mathbf{w}$ after each update to $\delta \mathbf{m}$---i.e.,
$\| \delta \mathbf{m} - g (\mathbf{z}, \mathbf{w}) \|_2^2 \simeq 0$, the
CNN has almost no effect in the next update for $\delta \mathbf{m}$. To
strike a balance between the number of updates to $\delta \mathbf{m}$
and $\mathbf{w}$, we choose to alternatingly take one gradient step for
$\delta \mathbf{m}$ and ten gradient steps for $\mathbf{w}$.

We use Devito \citep{devito-compiler, devito-api} to compute matrix-free
actions of the linearized Born scattering operator and its adjoint. By
integrating these operators into PyTorch, we are able to solve the
optimization problems~\ref{nlp-deep-prior} and~\ref{nlp-weak-deep-prior}
with automatic differentiation. We follow \citet{Lempitsky} for the CNN
architecture. We provide more details regarding to our implementation on
\href{https://github.com/slimgroup/Software.SEG2020/tree/master/siahkoohi2020SEGwdp}{GitHub}.

\section{Numerical experiments}\label{numerical-experiments}

\begin{figure}
\centering
\subfloat[\label{model-dm}]{\includegraphics[width=0.500\hsize]{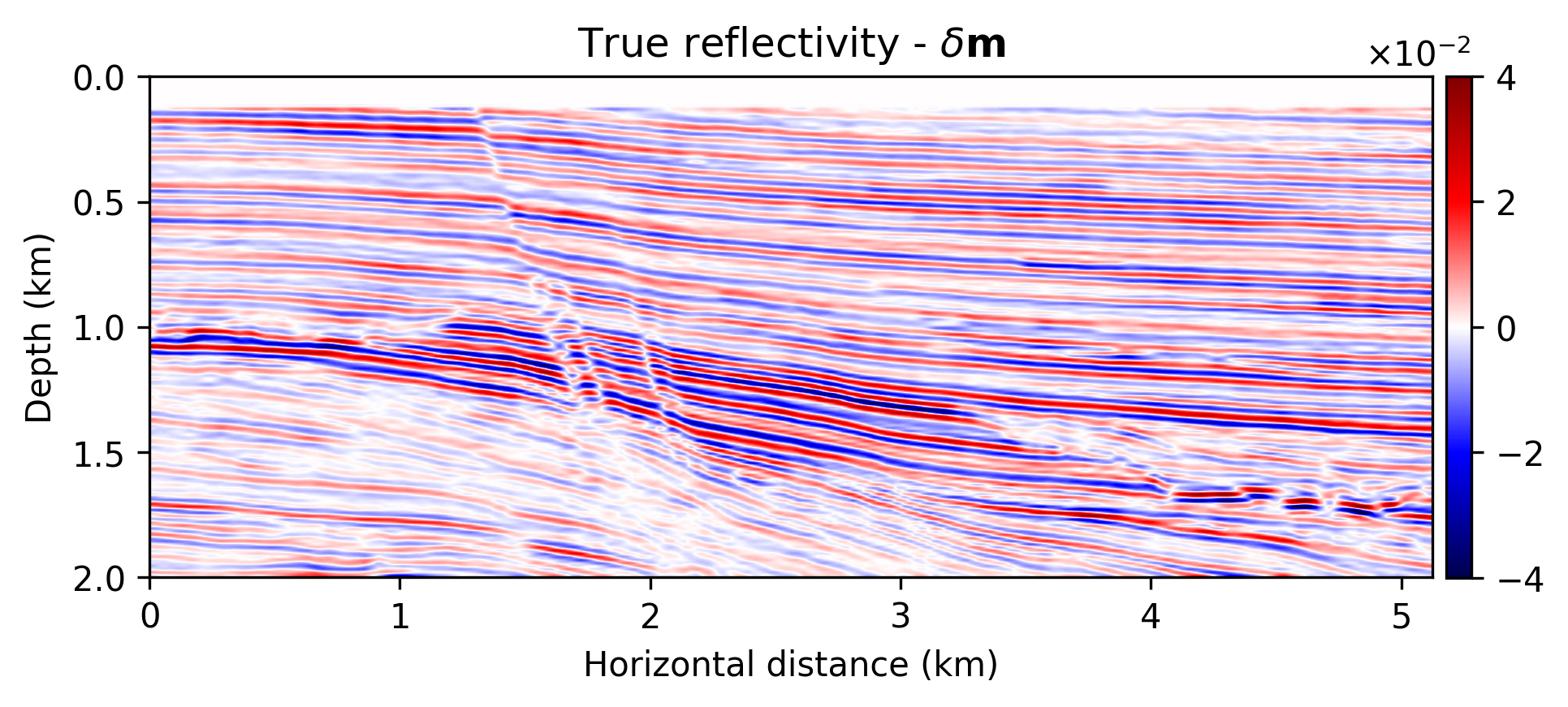}}
\\
\subfloat[\label{MLE_x}]{\includegraphics[width=0.500\hsize]{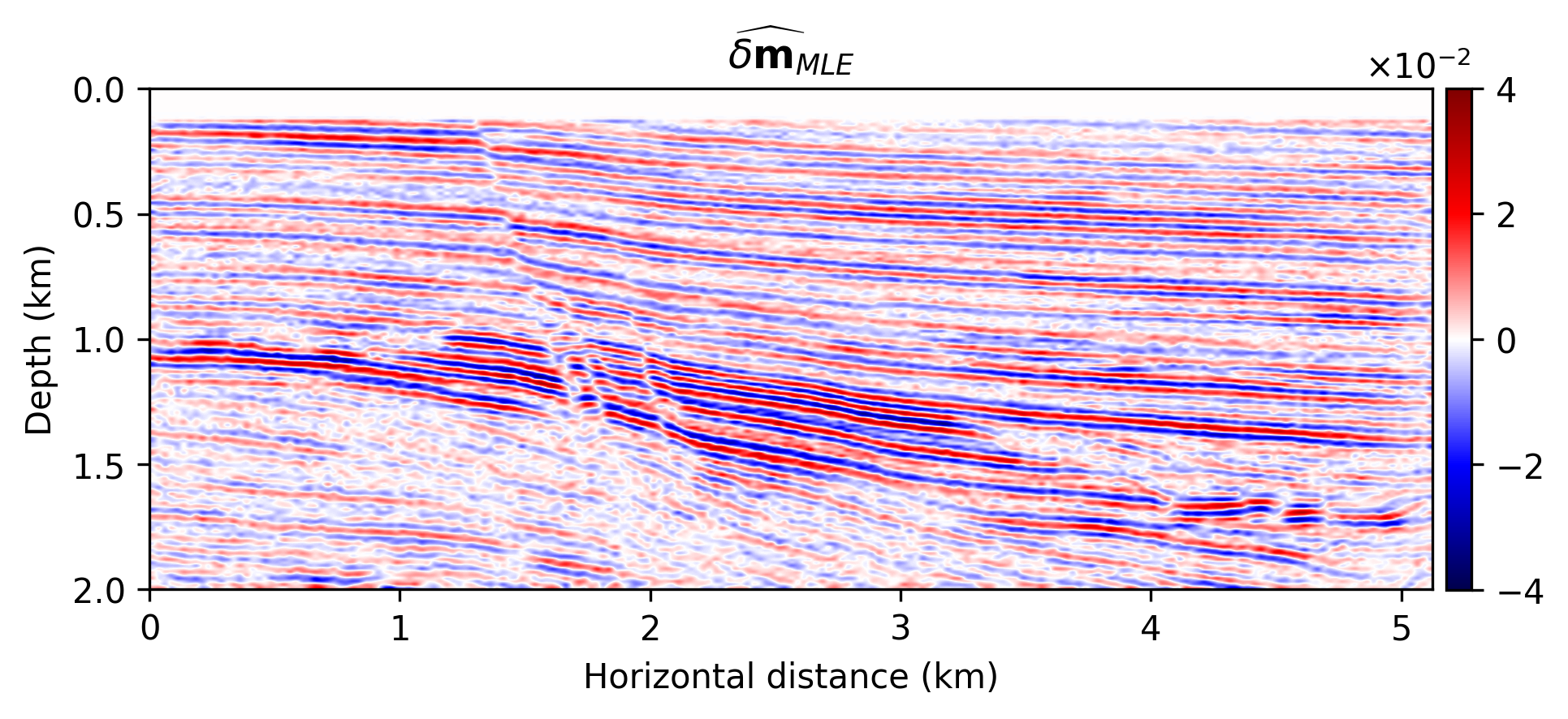}}
\subfloat[\label{MAP_w}]{\includegraphics[width=0.500\hsize]{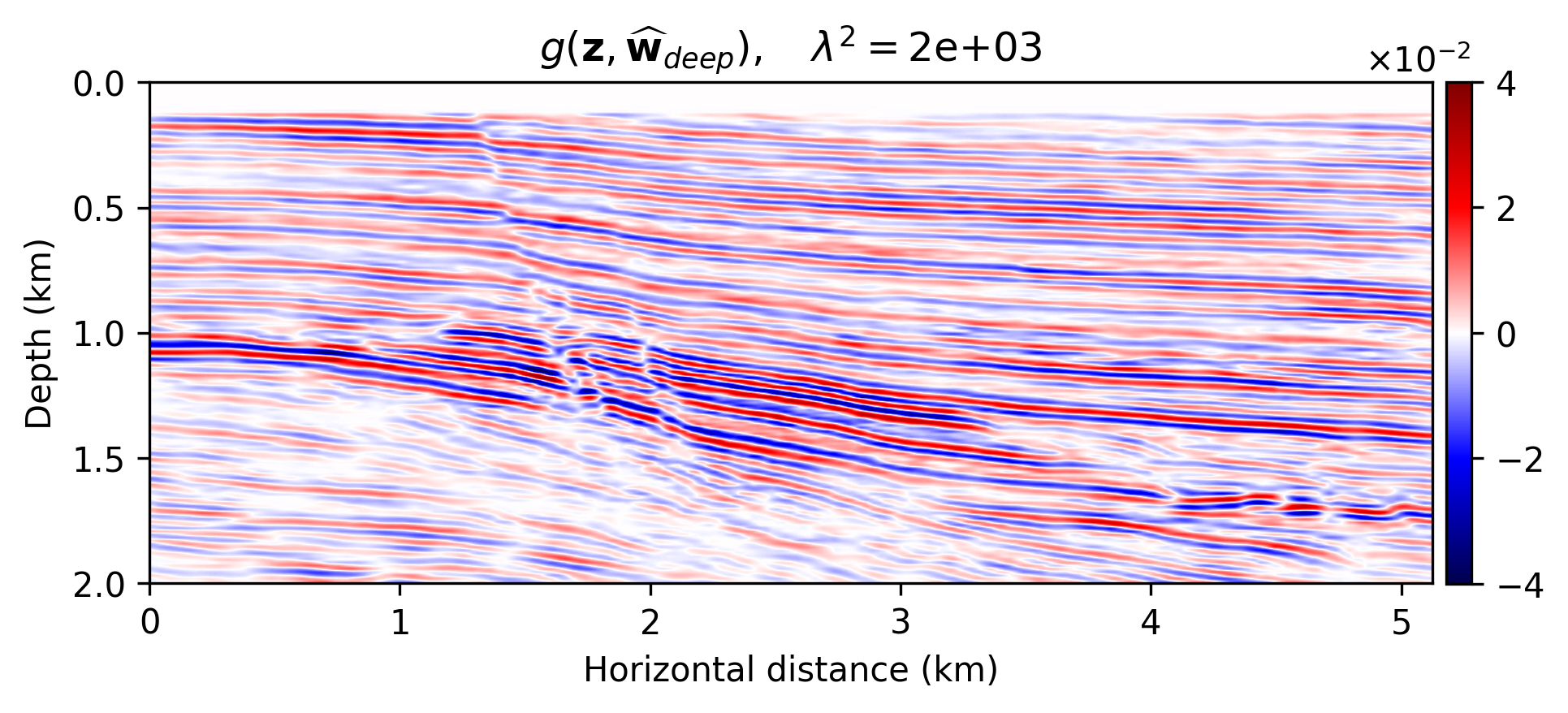}}
\\
\subfloat[\label{weak_MAP_w-1}]{\includegraphics[width=0.500\hsize]{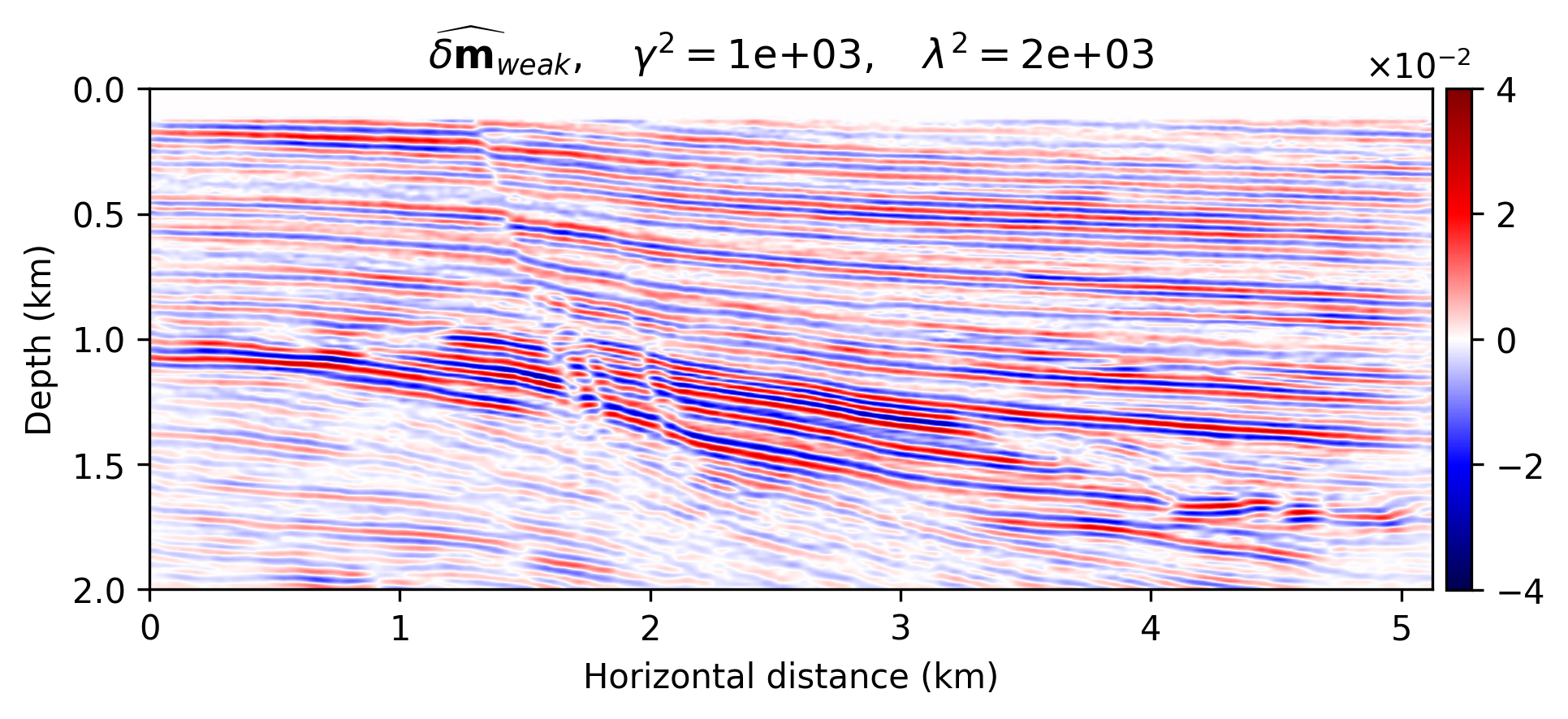}}
\subfloat[\label{weak_MAP_w-2}]{\includegraphics[width=0.500\hsize]{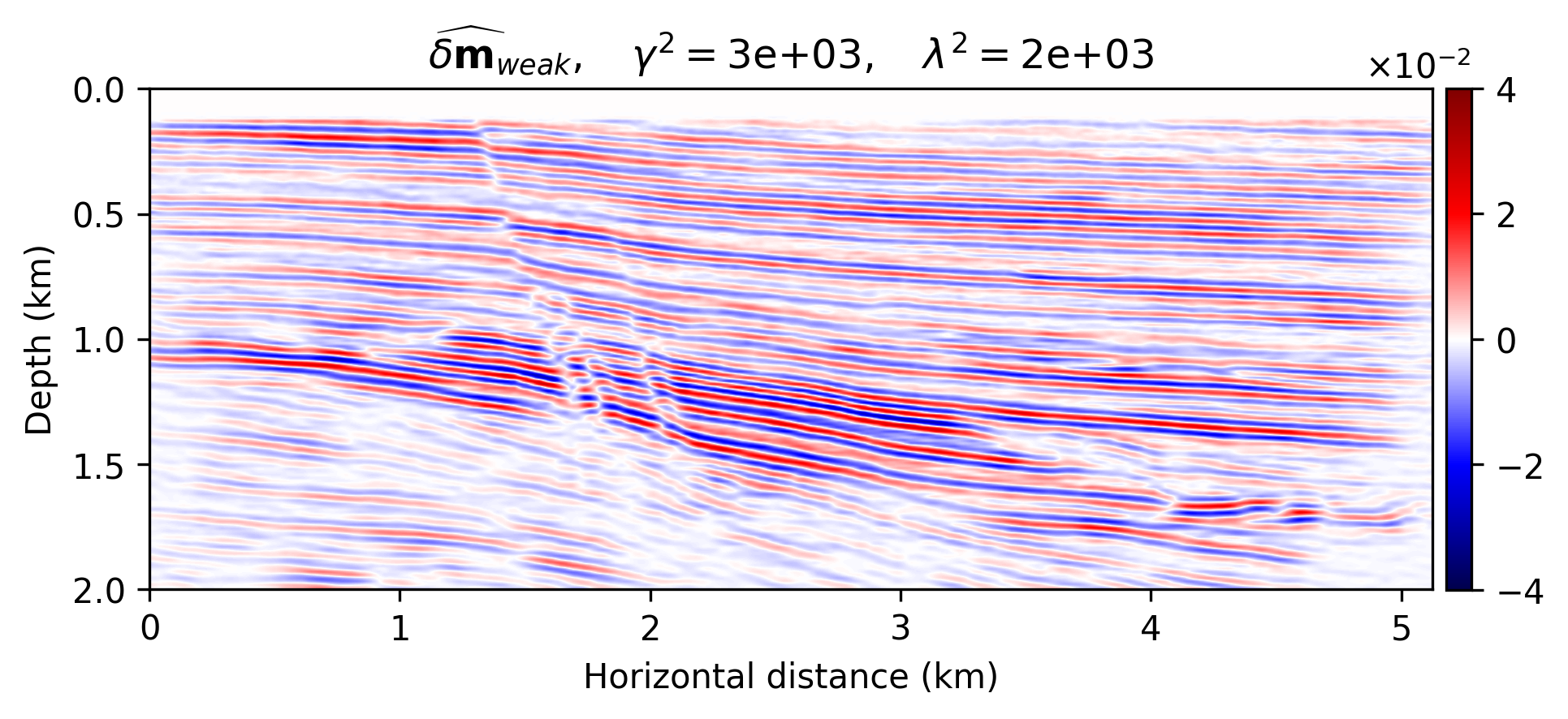}}
\caption{Imaging with the proposed method. a) True model. b)
$\widehat{\delta \mathbf{m}}_{\text{MLE}}$. c)
$g (\mathbf{z}, \widehat{\mathbf{w}}_{\text{deep}})$. d, e)
$\widehat{\delta \mathbf{m}}_{\text{weak}}$, with $\gamma = 10^3$ and
$3 \times 10^3$, respectively.}\label{results}
\end{figure}

We compare the seismic images obtained by solving
problems~\ref{nlp-deep-prior} and~\ref{nlp-weak-deep-prior}, when
applied to a ``quasi'' real field data example consisting of a 2D
portion of the Kirchoff time migrated
\href{https://wiki.seg.org/wiki/Parihaka-3D}{3D Parihaka} dataset (see
Figure~\ref{model-dm}). These imaging results are set as the ground
truth for the experiment here discussed. Synthetic data is obtained by
applying the linearized Born scattering operator to this ``true''
reflectivity image. The dataset includes $205$ shot records sampled with
a source spacing of $25\, \mathrm{m}$ and $1.5$ seconds recording time.
There are $410$ fixed receivers sampled at $12.5 \mathrm{m}$ spread
across the survey area. The source is a Ricker wavelet with a central
frequency of $30\,\mathrm{Hz}$. To demonstrate the regularization effect
of our method, we add a significant amount of noise to the shot records,
yielding a low signal-to-noise ratio of the ``observed'' data of
$-18.01\, \mathrm{dB}$. To limit the computational costs, we mix the
shot records according to normally distributed source encodings. By
conducting extensive parameter tuning, we set
$\lambda^2 = 2 \times 10^3$ (Equations~\ref{nlp-deep-prior}
and~\ref{nlp-weak-deep-prior}) throughout all experiments. We also set
$\sigma^2 = 0.01$ (Equations~\ref{nll}, \ref{nlp-deep-prior},
and~\ref{nlp-weak-deep-prior}), which is equal to the variance of the
measurement noise. To provide evidence regarding to the computational
feasibility of the weak deep prior formulation, we fix the number of
passes over the dataset---i.e., we use roughly twice the computational
cost of reverse-time migration (computational budget for large-scale
least-squares imaging) However, as mentioned before, the deep prior
formulation requires more iterations to generate a reasonable image. We
use $15$ passes over the dataset to solve
problem~\ref{nlp-deep-prior}---i.e., to compute
$\widehat{\mathbf{w}}_{\text{deep}}$. Note that taking one gradient step
for $\delta \mathbf{m}$ takes roughly $60$ times more time than one
update of $\mathbf{w}$, without GPU acceleration. Therefore, we neglect
the CNN weights update times in our comparisons.

The imaging results are included in Figure~\ref{results}.
Figure~\ref{model-dm} indicates the reflectivity that we have used to
generate linearized data. Figure~\ref{MLE_x} is the MLE image---i.e.,
conventional least-squares reverse-time migration,
$\widehat{\delta \mathbf{m}}_{\text{MLE}}$, obtained by minimizing
Equation~\ref{nll} with Adagrad for two passes over the
dataset.~Figures~\ref{MAP_w} shows the the deep-prior based image,
$g (\mathbf{z},\widehat{\mathbf{w}}_{\text{deep}})$, computed by running
RMSprop for $15$ passes over the dataset. Figures~\ref{weak_MAP_w-1}
and~\ref{weak_MAP_w-2} show the obtained results using the proposed
method by solving problem~\ref{nlp-weak-deep-prior} for two passes over
the dataset using values $\gamma = 10^3$ and $\gamma = 3 \times 10^3$,
respectively.

We make the following observations. As expected, Figure~\ref{MLE_x}
contains imaging artifacts since no prior regularization is in effect.
Although the deep prior has been successful in generating a realistic
result (Figures~\ref{nlp-deep-prior}), computing the solution required
$15$ passes over the source experiments, which is practically not
attainable for larger problems. The solution to the weak deep prior
imaging problem, with $\gamma = 10^3$
(Equation~\ref{nlp-weak-deep-prior}), generates a seismic image with
considerably less artifacts compared to MLE (compare Figures~\ref{MLE_x}
and~\ref{weak_MAP_w-1}), using the same number of wave-equation solves.
By comparing Figure~\ref{weak_MAP_w-1} with the image obtained by deep
prior based imaging (Figure~\ref{MAP_w}), we observe that the proposed
method is able to provide the benefits of deep prior while remaining
computationally feasible. However, Figure~\ref{weak_MAP_w-1} is slightly
less smooth compared to the true reflectivity (Figure~\ref{model-dm})
and deep prior based recovery (Figure~\ref{weak_MAP_w-1}). We can
increase the the deep prior penalty by increasing $\gamma$ to
$3 \times 10^3$ (Figure~\ref{weak_MAP_w-2}) to get an image with less
artifacts compared to Figure~\ref{weak_MAP_w-1}. The cost of heavier
penalty is amplitude underestimation compared to the deep-prior result
(Figure~\ref{MAP_w}).

\section{Conclusions}\label{conclusions}

The proposed method is an alternative to classical constrained
optimization, where handcrafted regularization is considered instead.
While practical and ubiquitous, the latter approach is based on
heavy-handed assumptions, which inevitably leaves a strong imprint on
the final result. Conversely, constraints by deep priors only requires
an uninformative Gaussian prior on the network weights. While deep
priors have been recently proven successful for many imaging problems, a
naive implementation for seismic imaging, which involves lengthy
wave-equation solvers, leads to a computationally expensive scheme. By
relaxing the deep prior, we decouple model and network updates when
optimizing, hence a relatively cheap training phase. As verified by our
numerical experiment, we are still able to resolve the imaging artifacts
present in conventional least-squares imaging when data is contaminated
by strong noise. Compared to reverse-time migration, the deep weak prior
approach requires twice its computational cost, an affordable
computational budget in large-scale imaging problems.

\bibliography{abstract}

\begin{thebibliography}{25}
\providecommand{\natexlab}[1]{#1}
\providecommand{\url}[1]{\texttt{#1}}
\expandafter\ifx\csname urlstyle\endcsname\relax
  \providecommand{\doi}[1]{doi: #1}\else
  \providecommand{\doi}{doi: \begingroup \urlstyle{rm}\Url}\fi

\bibitem[{Lempitsky} et~al.(2018){Lempitsky}, {Vedaldi}, and
  {Ulyanov}]{Lempitsky}
V.~{Lempitsky}, A.~{Vedaldi}, and D.~{Ulyanov}.
\newblock {D}eep {I}mage {P}rior.
\newblock In \emph{2018 IEEE/CVF Conference on Computer Vision and Pattern
  Recognition}, pages 9446--9454, June 2018.
\newblock \doi{10.1109/CVPR.2018.00984}.

\bibitem[Cheng et~al.(2019)Cheng, Gadelha, Maji, and Sheldon]{Cheng_2019_CVPR}
Zezhou Cheng, Matheus Gadelha, Subhransu Maji, and Daniel Sheldon.
\newblock {A Bayesian Perspective on the Deep Image Prior}.
\newblock In \emph{{The IEEE Conference on Computer Vision and Pattern
  Recognition (CVPR)}}, pages 5443--5451, June 2019.

\bibitem[Gadelha et~al.(2019)Gadelha, Wang, and Maji]{gadelha2019shape}
Matheus Gadelha, Rui Wang, and Subhransu Maji.
\newblock Shape reconstruction using differentiable projections and deep
  priors.
\newblock In \emph{Proceedings of the IEEE International Conference on Computer
  Vision}, pages 22--30, 2019.

\bibitem[Liu et~al.(2019)Liu, Fu, and Zhang]{liu2019deep}
Qun Liu, Lihua Fu, and Meng Zhang.
\newblock Deep-seismic-prior-based reconstruction of seismic data using
  convolutional neural networks.
\newblock \emph{arXiv preprint arXiv:1911.08784}, 2019.

\bibitem[Wu and McMechan(2019)]{wu2019parametric}
Yulang Wu and George~A McMechan.
\newblock Parametric convolutional neural network-domain full-waveform
  inversion.
\newblock \emph{{GEOPHYSICS}}, 84\penalty0 (6):\penalty0 R881--R896, 2019.
\newblock \doi{10.1190/geo2018-0224.1}.

\bibitem[Shi et~al.(2020)Shi, Wu, and Fomel]{shi2020deep}
Yunzhi Shi, Xinming Wu, and Sergey Fomel.
\newblock Deep learning parameterization for geophysical inverse problems.
\newblock In \emph{{SEG 2019 Workshop: Mathematical Geophysics: Traditional vs
  Learning, Beijing, China, 5-7 November 2019}}, pages 36--40. Society of
  Exploration Geophysicists, 2020.
\newblock \doi{10.1190/iwmg2019_09.1}.

\bibitem[Siahkoohi et~al.(2020)Siahkoohi, Rizzuti, and
  Herrmann]{siahkoohi2020EAGEdlb}
Ali Siahkoohi, Gabrio Rizzuti, and Felix~J. Herrmann.
\newblock A deep-learning based bayesian approach to seismic imaging and
  uncertainty quantification.
\newblock \emph{{82nd EAGE Conference and Exhibition 2020}}, 6 2020.
\newblock URL
  \url{https://slim.gatech.edu/Publications/Public/Submitted/2020/siahkoohi2020EAGEdlb/siahkoohi2020EAGEdlb.html}.

\bibitem[Herrmann et~al.(2019)Herrmann, Siahkoohi, and
  Rizzuti]{herrmann2019NIPSliwcuc}
Felix~J. Herrmann, Ali Siahkoohi, and Gabrio Rizzuti.
\newblock Learned imaging with constraints and uncertainty quantification.
\newblock In \emph{{Neural Information Processing Systems (NeurIPS) 2019 Deep
  Inverse Workshop}}, 12 2019.
\newblock URL \url{https://arxiv.org/pdf/1909.06473.pdf}.

\bibitem[Ovcharenko et~al.(2019)Ovcharenko, Kazei, Kalita, Peter, and
  Alkhalifah]{ovcharenko2019deep}
Oleg Ovcharenko, Vladimir Kazei, Mahesh Kalita, Daniel Peter, and Tariq~Ali
  Alkhalifah.
\newblock Deep learning for low-frequency extrapolation from multi-offset
  seismic data.
\newblock \emph{{GEOPHYSICS}}, 84\penalty0 (6):\penalty0 R989--R1001, 11 2019.
\newblock \doi{10.1190/geo2018-0884.1}.

\bibitem[Rizzuti et~al.(2019)Rizzuti, Siahkoohi, and
  Herrmann]{rizzuti2019EAGElis}
Gabrio Rizzuti, Ali Siahkoohi, and Felix~J. Herrmann.
\newblock {Learned iterative solvers for the Helmholtz equation}.
\newblock \emph{{81st EAGE Conference and Exhibition 2019}}, 2019.
\newblock ISSN 2214-4609.
\newblock \doi{10.3997/2214-4609.201901542}.
\newblock URL
  \url{https://www.slim.eos.ubc.ca/Publications/Private/Submitted/2019/rizzuti2019EAGElis/rizzuti2019EAGElis.pdf}.

\bibitem[Siahkoohi et~al.(2019{\natexlab{a}})Siahkoohi, Kumar, and
  Herrmann]{siahkoohi2019dlwr}
Ali Siahkoohi, Rajiv Kumar, and Felix~J. Herrmann.
\newblock {Deep-learning based ocean bottom seismic wavefield recovery}.
\newblock In \emph{{SEG Technical Program Expanded Abstracts 2019}}, pages
  2232--2237, 8 2019{\natexlab{a}}.
\newblock \doi{10.1190/segam2019-3216632.1}.

\bibitem[Siahkoohi et~al.(2019{\natexlab{b}})Siahkoohi, Louboutin, and
  Herrmann]{siahkoohi2019transfer}
Ali Siahkoohi, Mathias Louboutin, and Felix~J. Herrmann.
\newblock The importance of transfer learning in seismic modeling and imaging.
\newblock \emph{{GEOPHYSICS}}, 84\penalty0 (6):\penalty0 A47--A52, 11
  2019{\natexlab{b}}.
\newblock \doi{10.1190/geo2019-0056.1}.

\bibitem[Siahkoohi et~al.(2019{\natexlab{c}})Siahkoohi, Verschuur, and
  Herrmann]{siahkoohi2019srmedl}
Ali Siahkoohi, Dirk~J. Verschuur, and Felix~J. Herrmann.
\newblock {Surface-related multiple elimination with deep learning}.
\newblock In \emph{{SEG Technical Program Expanded Abstracts 2019}}, pages
  4629--4634, 8 2019{\natexlab{c}}.
\newblock \doi{10.1190/segam2019-3216723.1}.

\bibitem[Sun and Demanet(2019)]{sun2019extrapolated}
Hongyu Sun and Laurent Demanet.
\newblock Extrapolated full waveform inversion with convolutional neural
  networks.
\newblock In \emph{{SEG Technical Program Expanded Abstracts 2019}}, 2019.
\newblock \doi{10.1190/segam2019-3197987.1}.

\bibitem[Zhang and Alkhalifah(2019)]{zhang2019regularized}
{Zhen-Dong} Zhang and Tariq Alkhalifah.
\newblock Regularized elastic full waveform inversion using deep learning.
\newblock \emph{{GEOPHYSICS}}, 84\penalty0 (5):\penalty0 1SO--Z28, 9 2019.
\newblock \doi{10.1190/geo2018-0685.1}.

\bibitem[Veritas(2005)]{Veritas2005}
Veritas.
\newblock {Parihaka 3D Marine Seismic Survey - Acquisition and Processing
  Report}.
\newblock Technical Report New Zealand Petroleum Report 3460, New Zealand
  Petroleum \& Minerals, Wellington, 2005.

\bibitem[WesternGeco.(2012)]{WesternGeco2012}
WesternGeco.
\newblock {Parihaka 3D PSTM Final Processing Report}.
\newblock Technical Report New Zealand Petroleum Report 4582, New Zealand
  Petroleum \& Minerals, Wellington, 2012.

\bibitem[Tarantola(2005)]{tarantola2005inverse}
Albert Tarantola.
\newblock \emph{Inverse problem theory and methods for model parameter
  estimation}.
\newblock SIAM, 2005.
\newblock ISBN 978-0-89871-572-9.
\newblock \doi{10.1137/1.9780898717921}.

\bibitem[Welling and Teh(2011)]{welling2011bayesian}
Max Welling and Yee~Whye Teh.
\newblock {Bayesian Learning via Stochastic Gradient Langevin Dynamics}.
\newblock In \emph{{Proceedings of the 28th International Conference on
  International Conference on Machine Learning}}, {ICML’11}, pages 681--688,
  2011.

\bibitem[Peters et~al.(2019)Peters, Smithyman, and Herrmann]{peters2018pmf}
Bas Peters, Brendan~R Smithyman, and Felix~J Herrmann.
\newblock Projection methods and applications for seismic nonlinear inverse
  problems with multiple constraints.
\newblock \emph{{GEOPHYSICS}}, 84\penalty0 (2):\penalty0 R251--R269, 2019.
\newblock \doi{10.1190/geo2018-0192.1}.

\bibitem[Robbins(2007)]{Robbins2007ASA}
Herbert~E. Robbins.
\newblock A stochastic approximation method.
\newblock 2007.

\bibitem[Duchi et~al.(2011)Duchi, Hazan, and Singer]{adagrad}
John Duchi, Elad Hazan, and Yoram Singer.
\newblock Adaptive subgradient methods for online learning and stochastic
  optimization.
\newblock \emph{{Journal of Machine Learning Research}}, 12\penalty0
  (null):\penalty0 2121–2159, July 2011.
\newblock ISSN 1532-4435.
\newblock \doi{10.5555/1953048.2021068}.
\newblock URL \url{http://jmlr.org/papers/v12/duchi11a.html}.

\bibitem[Tieleman and Hinton(2012)]{rmsprop}
Tijmen Tieleman and Geoffrey Hinton.
\newblock Lecture 6.5-{RMS}prop: {D}ivide the gradient by a running average of
  its recent magnitude.
\newblock 2012.

\bibitem[{Luporini} et~al.(2018){Luporini}, {Lange}, {Louboutin}, {Kukreja},
  {H{\"u}ckelheim}, {Yount}, {Witte}, {Kelly}, {Herrmann}, and
  {Gorman}]{devito-compiler}
F.~{Luporini}, M.~{Lange}, M.~{Louboutin}, N.~{Kukreja}, J.~{H{\"u}ckelheim},
  C.~{Yount}, P.~{Witte}, P.~H.~J. {Kelly}, F.~J. {Herrmann}, and G.~J.
  {Gorman}.
\newblock Architecture and performance of devito, a system for automated
  stencil computation.
\newblock \emph{CoRR}, abs/1807.03032, jul 2018.
\newblock URL \url{http://arxiv.org/abs/1807.03032}.

\bibitem[Louboutin et~al.(2019)Louboutin, Lange, Luporini, Kukreja, Witte,
  Herrmann, Velesko, and Gorman]{devito-api}
M.~Louboutin, M.~Lange, F.~Luporini, N.~Kukreja, P.~A. Witte, F.~J. Herrmann,
  P.~Velesko, and G.~J. Gorman.
\newblock Devito (v3.1.0): an embedded domain-specific language for finite
  differences and geophysical exploration.
\newblock \emph{{Geoscientific Model Development}}, 12\penalty0 (3):\penalty0
  1165--1187, 2019.
\newblock \doi{10.5194/gmd-12-1165-2019}.
\newblock URL \url{https://www.geosci-model-dev.net/12/1165/2019/}.

\end{thebibliography}

\end{document}